\documentstyle[psfig,amstex]{mn}

\begin{document}

\title[The BBF]{The Bivariate Brightness Function of Galaxies and a 
Demonstration of the Impact of Surface Brightness Selection Effects on 
Luminosity Function Estimations.}
\author[Cross \& Driver]{\parbox[t]{\textwidth}
{Nicholas Cross, Simon P.\ Driver}
\vspace*{6pt} \\ 
School of Physics and Astronomy, North Haugh, St Andrews, Fife, 
KY16 9SS, United Kingdom \\
%{\rm(njgc, spd3@st-andrews.ac.uk)} \\
}

\maketitle

\begin{abstract}

In this paper we fit an analytic function to the Bivariate Brightness 
Distribution (BBD) of galaxies. It is a combination of the classical 
Schechter Function convolved with a Gaussian distribution in surface 
brightness: thus incorporating the luminosity-surface brightness correlation 
as seen in many recent datasets. We fit this function to 
a recent measurement of the BBD based on 45,000 galaxies from the {\it 
two-degree field Galaxy Redshift Survey} (Cross et al. 2001). The parameters 
for the best fit model are $\phi^*=(0.0206 \pm 0.0009)h^{3}$Mpc$^{-3}$, 
$M^*_{b_j}-5\log\,h=(-19.72 \pm 0.04)$ mag, $\alpha=-1.05 \pm 0.02$, 
$\beta_{\mu}=0.281 \pm 0.007$, $\mu^*_{e,b_j}=(22.45 \pm 0.01)$ mag 
arcsec$^{-2}$ and $\sigma_{\mu}=0.517 \pm 0.006$. $\phi^*$, $M^*_{b_j}$ and 
$\alpha$ equate to the conventional Schechter parameters. $\beta_{\mu}$ is the 
slope of the luminosity-surface brightness correlation, $\mu_{e,b_j}^*$ is the 
characteristic effective surface brightness at $M^*_{b_j}$ and $\sigma_{\mu}$ 
is the width of the Gaussian.  

Using a BBF we explore the impact of the limiting detection isophote on
classical measures of the galaxy luminosity distribution. We demonstrate that 
if isophotal magnitudes are used then errors of
$\Delta M^*_{b_j} \sim 0.62 $ mags,
$\Delta \phi^* \sim 26\%$ and
$\Delta \alpha \sim 0.04$ are likely for $\mu_{lim,b_j}=24.0$ mag 
arcsec$^{-2}$. If Gaussian corrected magnitudes are used these change to
$\Delta M^*_{b_j} \sim 0.38 $ mags,
$\Delta \phi^* \sim 11\%$ and
$\Delta \alpha < 0.01$ for $\mu_{lim,b_j}=24.0$ mag arcsec$^{-2}$.
Hence while the faint-end slope, $\alpha$, appears fairly robust to surface 
brightness issues, both the $M^*$ and $\phi^*$ values are highly dependent.
The range over which these parameters were seen to vary is fully consistent 
with the scatter in the published values, reproducing the range of observed
luminosity densities ($1.1<j_{b_j}<2.2\times10^8h\,L_{\odot}$Mpc$^{-3}$ see 
Cross et al. 2001). If total magnitudes are recovered then there is no change 
in the luminosity
function within the errors for $\mu_{lim,b_j}=24.0$ mag arcsec$^{-2}$. We
conclude that surface brightness selection effects are primarily responsible 
for this variation. After due consideration of these effects, we derive a 
value of $j_{b_j}=2.16\times10^8h\,L_{\odot}$Mpc$^{-3}$.

\end{abstract}


\section{Introduction}
For the past quarter of a century the luminosity distribution of galaxies 
has been represented by a Schechter function (Schechter 1976). This contains
three defining parameters (see Eqn.~\ref{eq:sch}) and from these parameters 
one can derive useful quantities such as the mean local luminosity density.
Over the past decade many measurements of the Schechter parameters have been 
made (e.g. Efstathiou, Ellis \& Peterson, 1988; Loveday et al. 1992; Marzke 
et al. 1994, 1998; Zucca et al., 1997). However when compared, the results 
from these recent surveys show a wide range in the derived Schechter 
parameters, Cross et al. (2001) and hence a wide range in the inferred 
luminosity density. 

The two most likely culprits for this variation are: cosmic 
variance and/or surface brightness selection bias (e.g. Disney 1976; 
Phillipps, Davies \& Disney 1990). In Cross et al. (2001) these two errors 
were quantified for a sample of 45,000 galaxies drawn from the two-degree 
field galaxy redshift survey (2dFGRS). The conclusion was that both of these 
issues are significant and can potentially lead to significant underestimates 
of the local luminosity density ($\sim$few \% for cosmic variance and 
$\sim$35\% for surface brightness biases). The surface brightness biases can 
be separated into two distinct parts; a surface brightness dependent 
photometric bias and a surface brightness dependent Malmquist bias. The 
biggest 
effect comes from the former, i.e. underestimating the total magnitudes 
(25\%). The remaining contribution comes from overestimating the volume over 
which a 
galaxy can be seen (10\%). Cross et al. (2001) advocated the construction of 
a Bivariate Brightness Distribution (BBD) to remove the surface 
brightness selection biases. The BBD is the space density of galaxies as a 
function of the absolute magnitude, $M$, AND the absolute surface brightness, 
$\mu$.
This process also revealed a number of noteworthy results in its own right:
Firstly a clear dearth of giant 
low-surface brightness galaxies; secondly a clear luminosity-surface brightness
correlation (spanning $-23 < M_{b_j} < -16$); and thirdly the general
rise in the space-density from giant to dwarf systems to the faint limit 
($M_{b_j} = -16$).

Some earlier attempts have been made to measure the BBD (e.g. Cho{\l}oniewski 
1985; van der Kruit 1987; Sodr\'e \& Lahav 1993). These were for small samples
of bright galaxies. The Cho{\l}oniewski sample contained 248 E/S0 galaxies
with $-22<M_B<-18$; the van der Kruit sample contained 51 galaxies with 
isophotal diameters $>2'$ at 26.5 mag arcsec$^{-2}$ (in the photographic 
IIIa-J band) and the Sodr\'e \& Lahav sample contained 529 galaxies with
isophotal diameters $>1'$ at 25.6-26.0 mag arcsec$^{-2}$ in the B-band. 
These are very strict limits, and only very intrinsically bright and large 
galaxies were well sampled. 

Recently three more extensive and independent measurements of the Bivariate 
Brightness Distribution have been made. Driver (1999) derived the BBD for a 
volume-limited sample drawn from the Hubble Deep Field, de Jong \& Lacey 
(2000)  studied a homogeneous sample of 1000 late-type spirals and Blanton et 
al. (2001) derived the BBD for a sample of 11,275 galaxies from the Sloan 
Digital Sky Survey. These three surveys plus Cross et al. (2001) all confirm 
the existence of the luminosity-surface brightness correlation,
demonstrating that {\it surface brightness selection biases
are luminosity dependent.} The luminosity-surface 
brightness correlation is measured to be $M_{b_j}=2.4\pm_{0.5}^{1.5}\mu_e-72.3
\pm_{32.9}^{11.0}$ from (Cross et al. 2001). This equates to $\mu_e \propto 
0.42 \pm_{0.26}^{0.10}M_{b_j}+30.2\pm_{5.1}^{2.1} $. From Hubble Deep Field 
data, Driver (1999) found a steeper gradient ($M_{F450W} \propto 1.5\mu_e$) as
did Ferguson \& Binggeli (1994) in the Virgo cluster ($M_B \propto 1.4\mu_o$).
The number density of galaxies is a maximum along this line, falling away at 
both higher and lower surface brightnesses. Given the luminosity-surface 
correlation it is hardly surprising if surveys with differing selection 
criterion recover widely ranging Schechter parameters.

A useful next step is to produce an analytical function to fit to these 
derived BBDs. In \S 2 we show such a function and fit it to the derived 
2dFGRS BBD. In \S 3 we discuss our fitting procedure and 
compare it to a similar estimation made by de Jong \& Lacey (2000). In \S4 we 
show how 
the luminosity density can be calculated from the BBF and compare it to the 
published values. Finally we explore the scope for error in Schechter function
estimators if surface brightness selection effects are ignored.

\section{A Bivariate Brightness Function}

The luminosity function as traditionally described by the Schechter Function 
(Schechter 1976), is shown below as Eqn.~\ref{eq:sch}:
 
\begin{equation}
\phi(M)dM=0.4\,\ln(10)\phi^*\,10^{0.4(M^*-M)(\alpha+1)}\,e^{-10^{0.4(M^*-M)}}
\label{eq:sch}
\end{equation}

This equation contains three parameters  M$^*$, $\phi^*$ \& $\alpha$ which
describe the ``characteristic magnitude'', the ``normalisation constant'' and 
the ``faint end slope'', respectively. Containing no surface brightness 
information this function provides a good fit to the space density of field
galaxies albeit over a fairly restricted range of luminosities 
($-22<M_{b_j}<-16$) in the field. However, the Schechter Function provides a 
poorer fit to the luminosity distribution in clusters (e.g. Driver et al. 
1994, Andreon, Cuillandre \& Pell\'o 2000) and when the population is 
subdivided according to spectral type (Madgwick et al. 2001). Current 
estimates constrain the three Schechter parameters, for field galaxies, to 
lie in the range: $-19.75 < M^{*} < -19.15; 0.013 < \phi^* < 0.027; -1.22 < 
\alpha < -0.7$ resulting in a pessimistic luminosity density range of: $1.1 < 
j_{b_j} < 3.2\times10^8h\,L_{\odot}$Mpc$^{-3}$.

Empirically Cross et al. (2001) find that the distribution of the galaxy 
population in surface brightness appears symmetrical about a ridge and can 
therefore be described by
a Gaussian or possibly a quadratic distribution. The ridge is described by 
$\mu = \beta_{\mu} M + C$, the luminosity-surface brightness correlation. 
Clearly a BBF needs to relate closely to the Schechter function in luminosity.
Hence by multiplying the classical Schechter function with a Gaussian in 
surface brightness we can construct the following BBF:

\begin{equation}
\begin{split}
\phi(M,\mu_e)=&\frac{0.4\ln(10)}{\sqrt{2\pi}\,\sigma_{\mu}}\phi_*\,
10^{0.4(M^*-M)(\alpha+1)}\,e^{-10^{0.4(M^*-M)}}\\
&\exp[-\frac{1}{2}(\frac{\mu_e-\mu_e^*-\beta_{\mu}\,(M-M_*)}{\sigma_{\mu}})^2] \\ 
\label{eq:func}
\end{split}
\end{equation}

\noindent where $\beta_{\mu}$ is the gradient of the luminosity-surface 
brightness correlation and $\sigma_{\mu}$ is the dispersion in the surface 
brightness. This function is identical to that presented by 
Cho{\l}oniewski (1985), and derived by Dalcanton, Spergel \& Summers (1997) 
and de Jong \& Lacey 
(1999a,b 2000) using the Fall \& Efstathiou (1980) disk-galaxy formation 
model. Note that the new term contains a 
normalisation coefficient $\frac{1}{\sqrt{2\pi}\sigma_{\mu}}$ ensuring that 
$\phi^*$, $M^*$ and $\alpha$ are identical to the traditional 
Schechter parameters.

\section{Fitting the BBF}

\begin{figure}
{\psfig{file=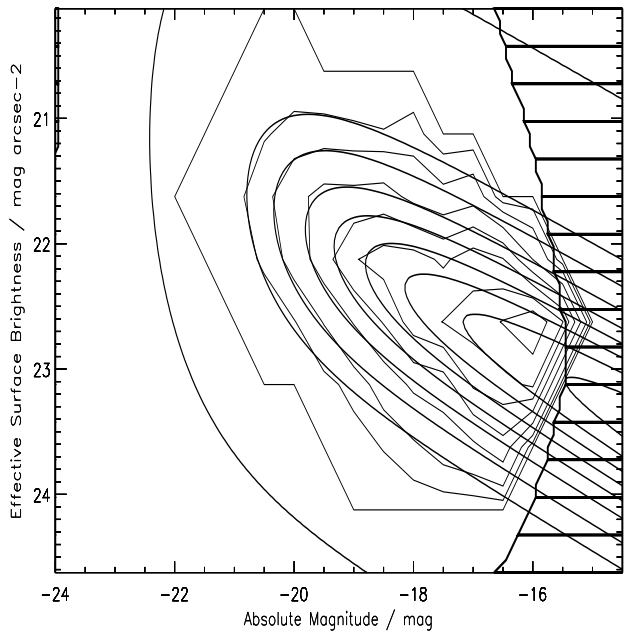,width=90mm,height=90mm}}
\caption{The thick lines show the BBF computed for the best fit parameters. 
The thin lines depict the 2dFGRS BBD (q.v. Cross et al. 2001 Fig. 10b). The 
contours are at $1.0\times10^{-7}$, $1.0\times10^{-3}$, $2.5\times10^{-3}$, 
$5.0\times10^{-3}$, $7.5\times10^{-3}$, $1.0\times10^{-2}$, $1.25\times
10^{-2}$, $1.5\times10^{-3}$, $1.75\times10^{-2}$, $2.0\times10^{-3}$ galaxies
Mpc$^{-3}$ mag$^{-1}$ (mag arcsec$^{-2}$). The shaded area shows the selection
boundary - see Cross et al. (2001) for details.\label{fig1}}
\end{figure}

We choose to fit the BBF to the data shown in Cross et al. (2001); this is 
the largest available data set. This BBD was derived from a subset of 45,000 
galaxies from the two-degree field galaxy redshift survey (see Cross et al. 
2001 for details), and is shown on Fig.~\ref{fig1} (thin contours). Also 
shown in Fig.~\ref{fig1} is the selection boundary derived from visibility 
theory (shaded region, see Appendix B of Cross et al. 2001). In the shaded 
region insufficient volume is surveyed to make any meaningful statement of 
the space-densities.

The BBF can be fitted to the BBD by minimising the $\chi^2$ of the model 
compared to the data.
The BBF is a non-linear six parameter equation and to find the minimum, we 
use the Levenberg-Marquardt Method (see Press et al. 1986). The data provided 
(see Cross et al. 2001) is binned as in Table C2 of Cross et al. (2001) using
those bins whose values are binned on a minimum of 25 galaxies.

The best fit parameters we derive are: 
$\phi^*=(0.0206 \pm 0.0009)h^3$Mpc$^{-3}$, $M^*-5\log\,h=(-19.72 \pm 0.04)$ 
mag, $\alpha=-1.05 \pm 0.02$, $\beta_{\mu}=0.281 \pm 0.007$, $\mu_e^*=(21.90 
\pm 0.01)$ mag arcsec$^{-2}$ and $\sigma_{\mu}=0.517 \pm 0.006$. All 
errors are $1\sigma$ errors. Fig.~\ref{fig1} shows the BBF for these 
parameters (thick lines) overlaid on the data (thin contour lines). The errors 
in the parameters were found using a Monte-Carlo simulation, that is the 
observed distribution was randomised within the quoted $1\sigma$ errors and 
the BBF fit re-derived. The final BBF fit yields a $\chi^{2}$ value of 164, 
for $\nu=49$, where $\nu$ is the no. of data points - no. of parameters. This 
gives a likelihood of $2.6\times10^{-14}$. 

Hence, although the BBF appears to describe the BBD, the fit is poor. 
It is important to understand where the differences are occurring. From 
Fig.~\ref{fig1} we see that the model fits the data well brightwards of 
$M=-18$, and less well in fainter bins. The errors become comparable to the 
space density faintwards of $M=-16$, so the main error in the fit occurs in 
the range: $-16>M>-18$. The data (Fig.~\ref{fig1}) show an upturn towards the 
faint end in this range whereas the Schechter function gradually flattens 
towards the faint end. Thus it is the
Schechter function part that does not describe the data well. Note that the 
BBF provides Schechter parameters comparable to the range from 
previous surveys. 

The model fits the data well in the surface brightness direction, implying 
that a Gaussian distribution is a good description of the space density as a
function of surface brightness, for a constant absolute magnitude. 
 
\subsection{Comparison with de Jong \& Lacey and Blanton}

Table~\ref{table1} compares our BBF with the de Jong \& Lacey (2000) BBF which 
was determined for Sb-Sdm galaxies only. As de Jong \& Lacey use 95\% 
confidence intervals for their errors, we have quoted $2\sigma$ errors
rather than $1\sigma$ errors for our values. We converted their half-light 
radii parameters to our effective surface brightness parameters. De Jong \& 
Lacey fit disks and exponential bulges 
to their data, taking into account inclination and internal extinction. In 
Appendix A, we estimate the uncertainty in our results due to not taking 
this into account and find that the error is $\sim0.1$ mag in $M*$ and 
$\sim0.55$ mag arcsec$^{-2}$ in $\mu_e^*$. We note that the de Jong \& 
Lacey (2000)  
parameters are their total galaxy parameters, not their disk-only galaxy 
parameters. In each case, these have been converted to $b_j$-band magnitudes 
using $B-I = 1.7$ mag from de Jong \& Lacey (2000), and $b_j=B-0.28(B-V)$
(Maddox, Efstathiou \& Sutherland 1990), using a value of $(B-V)=0.5$ for a 
late type spiral (Coleman, Wu \& Weedman, 1980, Driver et al. 1994). Finally 
we convert from $H_0 = 65$ km 
s$^{-1}$Mpc$^{-1}$ to $H_0 = 100$ km s$^{-1}$Mpc$^{-1}$. In addition, the de 
Jong \& Lacey sample has tighter selection criteria and more accurate CCD 
photometry ($\pm 0.05$ mag compared to $\pm 0.2$ mag), but it only includes 
late-type galaxies and has a redshift 
completeness of 80\% compared to $>90$\% for the 2dFGRS.
In spite of this, we find a similar spread in $\mu$ ($\sigma_{\mu}$ = $0.52$ 
q.v. $0.61$) and we find that the $\alpha$ values  of the two surveys are 
equal within the errors. The 2dFGRS has a brighter $M^*$ by 0.05 mag, and a 
brighter $\mu_e^*$ by 0.85 mag arcsec$^{-2}$. Taking 
into account the effects of bulges and inclination as mentioned above, the 
2dFGRS distribution has become 0.15 mag brighter than the de Jong \& Lacey 
distribution and has a brighter $\mu_e^*$ by 0.3 mag arcsec$^{-2}$. 
Considering that late-type galaxies tend to be fainter and lower surface 
brightness, these results appear fully consistent. 

Our value of $\beta_{\mu}$ 
can be converted to a luminosity-scale size gradient $\beta_{r_e}=-0.360
\pm 0.004$. Although this differs from the de Jong \& Lacey (2000)  value, 
interestingly, it agrees more closely with their theoretical prediction of 
$\beta_{r_e}=-\frac{1}{3}$ (see de Jong \& Lacey 2000). One possible reason
for the variation in $\beta_{\mu}$ may be a correlation between colour and
absolute magnitude. Blanton et al. (2001) find a strong correlation between
$(g^*-r^*)$ colour and $M_{r^*}$: brighter galaxies are redder, fainter 
galaxies 
are bluer. Making estimates of the colour-magnitude correlation from 
Fig. 13 of Blanton et al. (2001), we find that $M_{r^*}=-8.75\pm_2^4(g^*-r^*)
-14.88\pm_2^1$. Using a mean $b_j-r^*=1.1$ (calculated from Fukugita, Shimasaku
\& Ichikawa 1995 and Maddox, Efstathiou \& Sutherland 1990), and assuming 
that the additional colour term $\Delta(b_j-r^*)=\Delta(g^*-r^*)$ we calculate 
that the expected 
$\beta_{\mu,r^*}=0.36\pm0.23$. The value estimated from Fig. 10 of Blanton et
al. (2001) is $\beta_{\mu,r^*}=0.50\pm_{0.1}^{0.2}$. Thus the $r^*$ band 
luminosity-surface brightness correlation appears steeper than the $b_j$ band
luminosity-surface brightness correlation. A similar 
colour-magnitude correlation in $(b_j-I)$ could explain the discrepancy 
between our result and de Jong \& Lacey (2000) result.

The general good overall agreement between these substantially different 
surveys
is an important vindication of both results. Cross et al. (2001) has extended 
the de Jong \& Lacey conclusions to the full range of galaxy types with 
$M<-16$. However, the different values obtained for the luminosity surface 
brightness correlation may reflect a colour or morphologically dependent 
luminosity-surface brightness correlation. Blanton et al. (2001) seem 
to have found similar results, but have not fitted a function or tabulated 
their results.

\section{Calculating the Luminosity Density}
As for the Schechter function it is trivial to calculate the luminosity
density, j, by integrating the product of the BBF and the luminosity over the 
complete range of surface brightness and absolute magnitude.

\begin{equation}
\begin{split}
j&=\int^{\infty}_{-\infty}\int^{\infty}_{-\infty}L(M)\,\phi(M,\mu)dMd\mu \\
&=\phi^*\,L^*\,\Gamma(\alpha+2)=\phi^*\,10^{-0.4(M^*-M_{\odot})}\, 
\Gamma(\alpha+2) \\
\end{split}
\end{equation}

The solution is the same as the solution to the integral obtained from the
Schechter function. When calculated using the best fit parameters, the value of 
the luminosity density, $j_{b_j}=(2.16\pm0.14)\times10^8h\,L_{\odot}$
Mpc$^{-3}$\footnote{Cross et al. (2001) used
a value of $L_{\odot,b_j}=5.48$, rather than the correct value of 
$L_{\odot,b_j}=5.30$. The values for the luminosity density in this paper use
the correct value.}. 

In Blanton et al. (2001), the Sloan team get a $40\%$ higher value for the
luminosity density in the $b_j$ filter than the 2dFGRS team. Does this mean 
that 2dFGRS is missing some galaxies, or at least underestimating their fluxes?
For a start the values of $M^*$ are consistent, suggesting that both surveys
are correcting magnitudes properly. However the measurement of $\phi^*$ is 
over 30\% higher in Blanton et al. (2001). A more recent paper (Yasuda et
al. 2001) revises the SDSS luminosity density of Blanton et al., to 
$j_{b_j}=2.43\pm0.21\times10^{-2}h^3$Mpc$^{-3}$. This revision is based on a 
fit to the galaxy number counts, suggesting that the Blanton et al. region was 
overdense by 30\%. The revised $\phi^*$ value ($\phi^*=2.05 \pm 0.12 
^{+0.66}_{-0.28}\times10^{-2}h^3$Mpc$^{-3}$) is now consistent with our
measurement of $\phi^*=2.06 \pm 0.09\times10^{-2}h^3$Mpc$^{-3}$, in the $b_j$ 
band. 

Given this revised value of $\phi^*$, the Blanton result is still 10\% 
higher, but Blanton used a colour term $b_j=B-0.35(B-V)$, whereas the correct 
colour term for the APM, tested using EIS data is $b_j=B-0.28(B-V)$ (Peacock, 
private communication). When these two factors are taken into account, the 
luminosity densities are entirely consistent.
 
As demonstrated in Cross et al. (2001) the peak of the luminosity density lies
well inside the selection boundaries. When the function is integrated over the 
range $-24<M<-15.5$, $20.1<\mu_e<24.1$, the value obtained is 
$j_{b_j}=2.14\times10^8h\,L_{\odot}$Mpc$^{-3}$ as compared to the summed BBD 
which gives: $j_{b_j}=(2.11\pm0.20)\times10^8h\, L_{\odot}$Mpc$^{-3}$. These 
are the approximate selection boundaries, so the correspondence is excellent. 
Unless 
the distribution shows an upturn outside the selection boundary the 2dFGRS 
data have uncovered over 98\% of the local B-band luminosity density.  

\section{Exploring Surface Brightness Selection Effects}

In Cross et al. (2001), Fig. 1 showed the variation in the LF as measured 
from a number of recent redshift surveys. It was postulated that the 
large variation was due to 
surface brightness selection effects. In this section the variation of the 
luminosity function with the limiting detection isophote is explored and 
compared to the range of published luminosity functions. Throughout, an 
exponential surface brightness profile and an Einstein-de Sitter cosmology 
are assumed.

To calculate a derived luminosity function, we start with our fit to the BBF.
We take into account the overestimate in $\mu_e^*$ by 0.55 mag arcsec$^{-2}$,
derived in Appendix A, and use a value of $\mu_e^*=22.45$ mag arcsec$^{-2}$
[Note that in Cross et al. (2001) we used a less sophisticated method to 
estimate the offset caused by the bulge and arrived at a figure of 0.55 mag 
arcsec$^{-2}$].

\begin{figure}
{\psfig{file=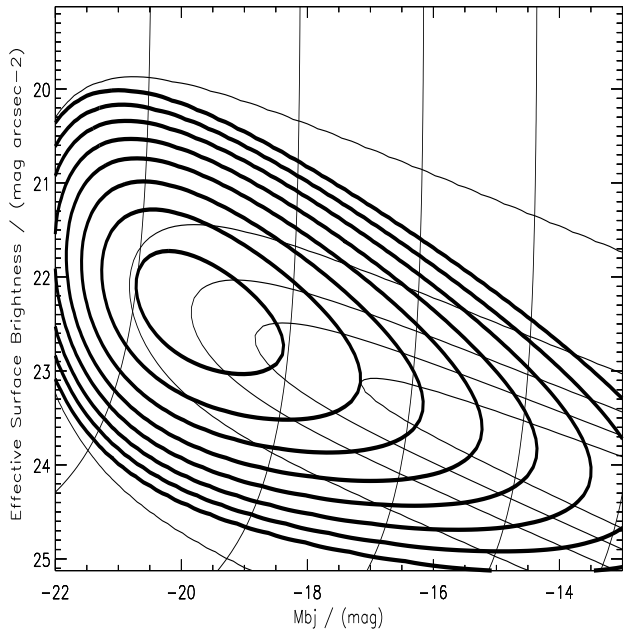,width=90mm,height=90mm}}
\caption{This figure shows 3 contour plots. The thin lines show the contours of
the visibility surface, in Mpc$^{3}$bin$^{-1}$. Each line is a decade apart. 
The five lines shown range from 10$^7$ to 10$^3$, from the bright end to the 
faint end respectively. This visibility surface has a detection threshold of 
26.0 mag arcsec$^{-2}$ and the limits are in isophotal magnitudes (see \S 5). 
The medium thickness lines show the BBF from Fig.~\ref{fig1}, offset in 
surface brightness by 0.55 mag arcsec$^{-2}$. The thick lines 
show the number of galaxies detected in each bin. The contours levels are at 
1.0, 3.2, 10, 32, 100, 320, 1000, 3200 and 10000 galaxies mag$^{-1}$ (mag 
arcsec$^{-2}$)$^{-1}$. 
\label{fig:ph_v_n}}
\end{figure}

\begin{figure}
{\psfig{file=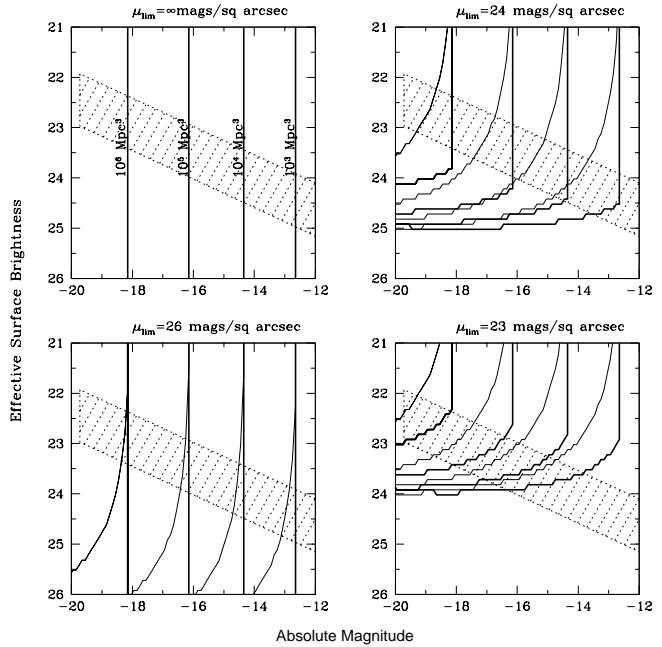,width=90mm,height=90mm}}
\caption{This plot shows contours of equal volume for different selection 
functions. The top, left plot shows the selection function for an infinite
threshold. The selection lines have no surface brightness dependence.
The other plots show isophotal (thin lines) and total (thick lines) selection 
functions, for the detection thresholds 26, 24, \& 23 mag arcsec$^{-2}$. 
As one goes to brighter thresholds, the volume becomes a stronger function of
surface brightness. Thus the mean volume at any $M$ decreases, and a volume 
correction or estimator which is only magnitude dependent becomes more biased.
The shaded parallelograms represent the number density distribution, in terms 
of the luminosity-surface brightness correlation, and the width of the 
surface brightness distribution. Where the contour lines are vertical, an
estimator with magnitude dependence only is unbiased, elsewhere it is biased.
Where the contour line crosses the distribution, the luminosity function can 
be recovered using an estimator which takes into account surface brightness. 
Where the distribution is missed, there is input catalogue incompleteness, and
so there is not enough information to recover the luminosity function.
When total magnitudes are used: at 26 mag 
arcsec$^{-2}$, there is no surface brightness dependency within the shaded 
region for $M<-12$, so a magnitude-only will give an unbiased luminosity 
function; at 24 mag arcsec$^{-2}$, it will be unbiased for $M<-14.3$, but an 
estimator with surface brightness built in, such as the method in Cross et 
al. (2001), will be unbiased $M<-12$; at 23 mag arcsec$^{-2}$, the magnitude
only estimator will be biased at all magnitudes, but a surface brightness 
estimator will recover the luminosity function for $M<-16$. Using isophotal
magnitudes, all luminosity function estimators will be biased if they are
only magnitude dependent.    
\label{fig:V_M}}
\end{figure}

We multiply our updated BBF by a visibility volume (Cross et al. 2001) to 
construct an apparent observed number in $M$ and $\mu$ 
(see Fig.~\ref{fig:ph_v_n}). The parameters adopted to derive the visibility 
surface are: $m_{faint}=20.0$ mag, $m_{bright}=14.0$ mag, $d_{min}=2.0''$, 
$d_{max}=250.0''$, $z_{max}=0.5$, $z_{min}=0$. The solid angle used was 
300$\Box^{\circ}$. These parameters are typical of the observed ranges
for the most recent surveys. The only parameter allowed to vary is the
detection isophote which took the values 26, 25, 24, 23.5 and 23 mag 
arcsec$^{-2}$. Fig.~\ref{fig:V_M} shows the complex Malmquist
bias for $\mu_{lim}=\infty, 26, 24, 23$ mag arcsec$^{-2}$ (top to bottom). The 
shaded region shows the approximate location of the galaxy population. The 
lines are contours of constant volume for that $\mu_{lim}$. As the limiting 
isophote becomes brighter, the surface brightness dependency of the Malmquist 
bias increases.

\begin{figure}
{\psfig{file=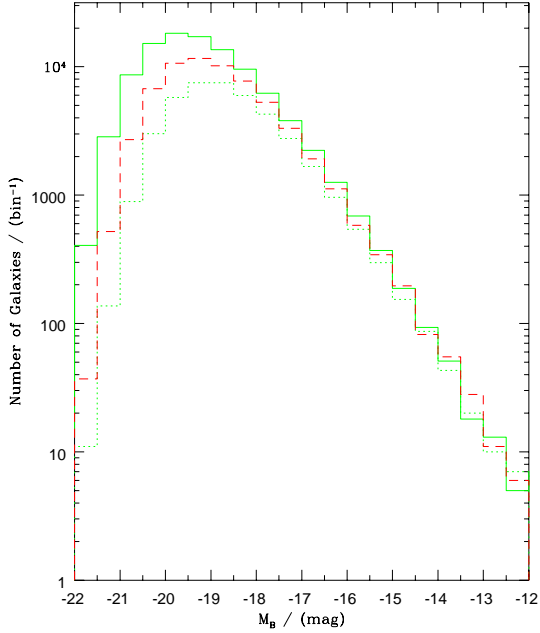,width=90mm,height=90mm}}
\caption{This is a plot of the number of galaxies detected as a function
of absolute magnitude, for a limiting isophote of 24.0 mag arcsec$^{-2}$. The 
solid line shows the plot for total magnitude, the dashed line shows the 
corresponding corrected magnitudes and the the dotted line shows the
corresponding isophotal magnitude. The difference between each line is a 
combination of an offset in magnitude and fewer galaxies being detected going
from total to isophotal magnitudes. The peak of the total magnitudes is 
$\sim0.3$ mag brighter than the peak in the corrected magnitudes which is 
$\sim0.35$ mag brighter than the peak in the isophotal magnitudes. 
\label{fig:N_M}}
\end{figure}

\begin{figure*}
{\psfig{file=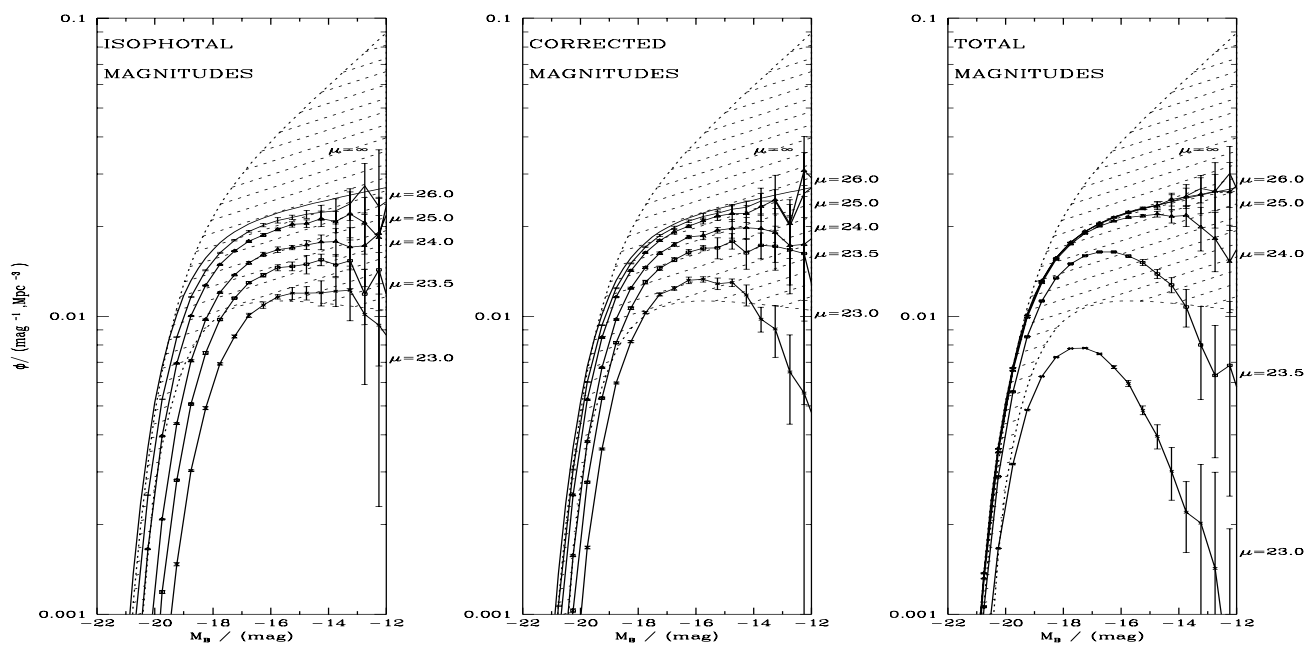,width=180mm,height=100mm}}
\caption{The variation of LFs with $\mu_{lim}$ for isophotal (top), corrected
(middle) and total (bottom) magnitudes. $\mu_{lim}$ varies from 26 mag 
arcsec$^{-2}$ to 23 mag arcsec$^{-2}$. The shaded region shows the variation 
of recent surveys from Sloan (with the new $\phi^*$) to APM. The LCRS is not 
shown as it has an additional surface brightness constraint that 
significantly reduces the faint end.
\label{fig:lfs}}
\end{figure*}

Given the observed distribution, each galaxy within each bin is then randomly 
assigned a volume out to the maximum derived from visibility theory. 
This volume is converted 
to a redshift assuming an Einstein-de Sitter cosmology and a standard 
k-correction. We assume that the number density does not vary within the
bin as a function of redshift, i.e. there is no evolution and no clustering. 
The exact absolute magnitude and surface brightness value is then randomly 
assigned within each bin, (this assumes that the number density does not vary 
within the bin as a function of $M$ or $\mu_e$). Around the $M^*$ point 
particularly, this assumption fails, but a Monte Carlo simulation done at
higher resolution finds no significant differences. The numbers in Table 2, 
and the plots in in Fig~\ref{fig:ph_v_n}-~\ref{fig:lfs} were produced from this
Monte Carlo simulation. 

The net result is a magnitude-limited sample with objects randomly 
distributed within their allowed volume. We now calculate each galaxy's 
isophotal magnitude, a Gaussian\footnote{One simple and popular method to 
correct for light lost is the Gaussian correction employed by Maddox, 
Efstathiou \& Sutherland (1990) on the APM, and used as a correction in the 
Source Extractor code (Bertin \& Arnouts 1996). It works by fitting a 
Gaussian with central surface brightness, $\mu_o$, and standard deviation, 
$\sigma$, to the light profile of the galaxy, such that the isophotal radius 
of the Gaussian matches the isophotal radius of the galaxy and the isophotal 
magnitude of the Gaussian is equivalent to the isophotal magnitude of the 
galaxy. The Gaussian corrected magnitude is then the total flux under the 
Gaussian. This works well for compact objects such as stars and small 
angular scale size galaxies where the seeing dominates the profile.} 
corrected magnitude and their total magnitude and sum the final 
number distribution according to absolute magnitude. This is plotted in 
Fig~\ref{fig:N_M} for total, corrected and isophotal absolute magnitudes.

We then reconstruct the luminosity function using a $1/V_{Max}$ prescription 
(as our simulations contain no clustering this should be an optimal estimator).
Fig.~\ref{fig:lfs} shows the recovered luminosity functions. The LFs of 
Fig.~\ref{fig:lfs} demonstrate the impact of surface brightness selection
as they are all drawn from the same BBF; the only difference is the 
limiting isophote and the choice of magnitude measurement. The range of 
published values is shown as the shaded area (excluding the LCRS Lin et al. 
1996). Also shown is the limit solution, for our model BBF. The left panel 
assumes isophotal magnitudes are measured, the central panel assumes Gaussian
corrected magnitudes were used and the right panel assumes some procedure has 
been implemented to recover the total magnitudes. The results are also 
tabulated in Table~\ref{table2}.

\subsection*{Isophotal magnitudes}
If isophotal magnitudes are adopted and the surface brightness limit is 
bright, the luminosities of galaxies are severely underestimated.
Thus both the number
density and the $M^*$ value are severely underestimated 
(see Table~\ref{table2}). The variation in $\phi^*$ is upto 50\% and in 
$M^*$  upto 1.0 mags. This tallies well with the range of Schechter values 
recovered (see \S 2) over the range tested ($23<\mu_{lim}<26$). To
some extent is it surprising that $\phi^*$ is not more drastically effected;
this is because the observed distribution of galaxies is skewed towards the 
faint end, see Fig.~\ref{fig:N_M}. As a simple $1/V_{max}$ correction or 
maximum likelihood estimator based on the isophotal magnitudes alone does not 
take into account surface brightness issues, especially light loss, a smaller 
volume is calculated than for total magnitudes, leading to 
an overestimate of the number density, see Fig~\ref{fig:V_M}. This is 
tempered by a lower number density at brighter absolute magnitudes.

Perhaps most surprising is the robustness of the faint end slope 
whose value is recovered correctly regardless of the isophote.

\subsection*{Corrected magnitudes}

Most surveys attempt to correct their isophotal magnitudes to total magnitudes.
We used a Gaussian correction as described above. 
Fig.~\ref{fig:lfs} and Table~\ref{table2} demonstrate that corrected 
magnitudes recover 63\% of the luminosity density at 24 mag arcsec$^{-2}$ 
compared to the 43\% that isophotal magnitudes recover and the 98\% that 
total magnitudes
recover. As with isophotal magnitudes, corrected magnitudes give a luminosity
function biased at all values of $M$, although the bias has been significantly
reduced.

\subsection*{Total magnitudes}
If some method is employed to correct the galaxies to total magnitudes 
(e.g. Kron magnitudes or Petrosian magnitudes) we find that the parameters are
very robust for $\mu_{lim} \geq 24$ mag arcsec$^{-2}$. However, at 
$\mu_{lim}=23$ mag arcsec$^{-2}$ the number density is underestimated 
throughout the distribution. Fig~\ref{fig:V_M} illustrates why this occurs.

The volume has almost no surface brightness dependency for the thresholds 
$24<\mu_{lim}<26$ provided $M<-14$, but it has significant surface brightness 
dependency for $-22<M<-14$ at $\mu_{lim}=23$. The bright absolute magnitudes 
are affected because of cosmological dimming and the K-correction. Galaxies 
at the maximum redshift of $z=0.5$ will have an apparent surface brightness
3 mag arcsec$^{-2}$ fainter than their intrinsic surface brightness (1.7 mag
arcsec$^{-2}$ due to cosmological dimming and 1.3 mag arcsec$^{-2}$ due to 
the K-correction, where $K(z)=2.5z$). Even 
galaxies at $z=0.25$ will be fainter by 1.6 mag arcsec$^{-2}$. Thus a galaxy
with central surface brightness of 21.5 mag arcsec$^{-2}$ ($\mu_e=22.6$ mag
arcsec$^{-2}$) and $z=0.25$ will not be detected with a threshold of $23$ mag
arcsec$^{-2}$. 

Recovering total magnitudes beforehand will give good estimators for the 
luminosity function, provided that significant numbers of galaxies are not 
missing. This is a particular problem if your maximum redshift is very high.
However, if galaxies are missing because of cosmological effects, rather than
being too intrinsically dim even at $z=0$, the number density can be recovered
using a surface brightness dependent volume correction as discussed in Cross
et al. (2001).

\subsection*{Overall Effects}

Overall the variations recovered in $M^*$, $\phi^*$ and $\alpha$ between 
simulated surveys with limits of $24 < \mu < 26$ (i.e., comparable to 
existing surveys) are $-19.74 < M^* < -19.08$, $0.020 < \phi^* < 0.017$ and 
$-1.07 < \alpha < -1.05$. Fig.~\ref{fig:lfs} shows that the observed 
variation in Schechter function 
parameters has been recovered at the bright end for $24<\mu_{lim}<26$ when 
either isophotal or Gaussian corrected magnitudes are used. However, 
Fig.~\ref{fig:lfs} demonstrates that the observed variation in the
faint end slope has not been recovered and is testament to the fact that 
surface brightness selection effects {\it do not} reproduce all the 
variation seen in the faint-end slope. This supports the suggestion in Cross 
et al. (2001) that the faint-end slope depends more critically on the 
clustering correction than surface brightness issues. 

Blanton et al. (2001) find a 0.08 change in the 
faint end slope going from $\mu_{e,r^*}=23.5$ to $\mu_{e,r^*}=24.5$, with no 
change at the bright end. Using a $(b_j-r^*)=1.1$, this leads to a $b_j$-band 
isophote of $\mu_{lim}=23.5$ for $\mu_{e,r^*}=23.5$. The change in the faint 
end can be compared to the changes seen in Table 2, for total magnitudes. The 
faint end slope, $\alpha\sim-1.037$ for $\mu_{lim}=23.5$ and $\alpha\sim-1.058$
for $\mu_{lim}=24.5$. This gives a 0.021 change over a similar interval, lower
than the Sloan result. However, Sloan has a red selected sample, which gives a 
steeper luminosity-surface brightness correlation (see \S 3.1). This could 
account for a greater change in $\alpha$.

In Cross et al. (2001) we take an isophotally selected sample and apply 
corrections in surface brightness as well as absolute magnitude. These 
corrections imply that we will not underestimate $M^*$ or $\phi^*$. SDSS 
calculated Petrosian magnitudes before selecting their sample. Petrosian 
magnitudes are aperture magnitudes and therefore do not show such a 
pronounced variation with redshift as isophotal magnitudes. 
Thus the sample is selected from pseudo-total limits.

For deep isophotes, the volume correction has virtually no surface brightness 
dependence, as shown in Fig~\ref{fig:lfs}, so the number density can be 
calculated at the bright end trivially and is only underestimated at the 
faint end where some galaxies have too a low surface brightness to get into 
the sample. Using either of the techniques outlined in the previous paragraph
(Cross et al. 2001, Blanton et al. 2001) should give accurate values of 
$M^*_{b_j}-5\log\,h$ and $\phi^*$, $-19.75 \pm 0.05$ mag, $(2.02 
\pm 0.02)\times10^{-2}h^3$Mpc$^{-3}$ for 2dFGRS 
(Cross et al. 2001) and $-19.70 \pm 0.04$ mag, 
$(2.05 \pm 0.12)\times10^{-2}h^3$Mpc$^{-3}$ for SDSS (Blanton et al. 2001, 
Yasuda et al. 2001). 
The caveat is that no correction to
total magnitudes is perfect, and the corrected magnitudes will tend to have
some surface brightness dependency as is demonstrated by the Gaussian 
corrected luminosity function. Even when isophotal magnitudes have been 
corrected to pseudo-total magnitudes it is better to use a $1/V_{max}$ or 
maximum likelihood estimator which is a function of both $M$ and $\mu$.

While a good understanding of visibility theory will account for galaxies 
within the surface brightness limits, galaxies with very low central surface
brightness, but bright total magnitudes can be missed. These are one source 
of mismatches between the estimates of $\alpha$ in different surveys. Another
reason for different estimates is inhomogeneities in the space density of 
galaxies. Surveys looking at different parts of the sky will encounter 
variations in the Large Scale Structure. Dwarf galaxies are seen over a 
smaller volume, so they can have large clustering corrections. Differences in 
clustering corrections between different surveys will tend to bias $\alpha$ 
rather than $\phi^*$.

\section{Conclusions}
We have presented a fitting function for the Bivariate Brightness Distribution.
This takes a similar form to a Schechter function in luminosity coupled with a 
Gaussian distribution in surface brightness. The Bivariate Brightness Function
was fitted to the recent results from Cross et al. (2001) who constructed a
BBD for a sample of 45,000 galaxies from the 2dFGRS. The BBF fits the data 
well at the bright end, but poorly at the faint end. 

We compare the parameters of the fit to the de Jong \& Lacey (2000) results
for late-type spirals. While our results broadly agree there are differences
which may provide clues toward understanding formation and evolution tracks 
of different galaxy types. 

The BBF can be integrated to yield a total luminosity density of 
$j_{b_j}=2.16\pm0.14\times10^8h\,L_{\odot}$Mpc$^{-3}$. This agrees with the 
luminosity density calculated from the data and demonstrates 
that unless the BBD shows sub-structure or a dramatic upturn beyond the 
selection boundaries the majority of the luminosity density in the local 
Universe has now been detected.

This paper has dealt with the biases that occur if
you start with an isophotally selected sample and do not apply any 
corrections, or apply a light loss correction without properly considering 
the volume correction. In both cases the Schechter parameters can be biased, 
and the luminosity density underestimated.

Using a BBF we explore the impact of the limiting detection isophote on
classical measures of the Schechter function. We demonstrate that 
if isophotal magnitudes are used then errors of
$\Delta M^*_{b_j} \sim 0.62 $ mags,
$\Delta \phi^* \sim 26\%$ and
$\Delta \alpha \sim 0.04$ are likely at $\mu_{lim,b_j}=24.0$ mag arcsec$^{-2}$.
If Gaussian corrected magnitudes are used these change to
$\Delta M^*_{b_j} \sim 0.38 $ mags,
$\Delta \phi^* \sim 11\%$ and
$\Delta \alpha < 0.01$ are likely at $\mu_{lim,b_j}=24.0$ mag arcsec$^{-2}$.
If total magnitudes can be recovered then the observed luminosity
function will be correct within the errors provided $\mu_{lim,b_j}>24.0$ mag 
arcsec$^{-2}$.
Hence while the faint-end slope, $\alpha$, appears fairly robust to surface 
brightness issues both the $M^*$ point and $\phi^*$ are highly dependent.
The range over which these parameters vary is fully consistent with the 
scatter in the published values which come from a variety of surveys with 
differing selection criterion. These parameters produce a range in the 
luminosity density, $j_{b_j}$, of $0.9 < j_{b_j} < 
2.2\times10^8h\,L_{\odot}$Mpc$^{-3}$ again agreeing well with the range of 
published values ($1.1-2.2\times10^8h\,L_{\odot}$Mpc$^{-3}$ see Cross et al. 
2001).

When selection effects are taken into account properly, the luminosity 
functions of recent surveys agree very well. The 2dFGRS and SDSS luminosity 
functions give the same $M^*$ and $\phi^*$ values, but disagree on the values
of $\alpha$. The differences in $\alpha$ are likely to be due to a 
combination of input catalogue incompleteness, filter used and different 
clustering corrections. 

\S 5 suggests that if a deep isophote is used and total magnitudes are 
recovered, then traditional magnitude dependent estimators can be used. 
However, future
work at higher redshifts or low redshift work on extremely faint galaxies
will include many galaxies close to the limits of the detection threshold, 
where surface brightness dependencies are strong. The BBD provides a framework 
that allows one to determine whether a bias is present (see Fig.~\ref{fig:V_M})
and to correct for it. In addition, it produces parameters such as 
$\beta_{\mu}$ 
and $\sigma_{\mu}$ which could be used to place constraints on galaxy 
formation and evolution models.

Our conclusion is that after a quarter of a century we need to upgrade our
representation of the space-density of galaxies to now include surface 
brightness. The BBD and BBF provide an excellent starting point and should lead
to more reliable and consistent measurements of the local luminosity density
as well as providing new constraints on galaxy formation models.

\begin{table}
\caption{Bivariate Brightness Function Parameters \label{table1}}
\begin{tabular}{|l|c|c|} \hline
Parameter & 2dFGRS & dJ \& L (2000)  \\ \hline \hline
$M^*_{b_j}-5\log\,h$ / mag & $-19.72 \pm 0.08$ & $-19.67 \pm 0.17$ \\ \hline
$\alpha$ & $-1.05 \pm 0.04$ & $-0.93 \pm 0.10$ \\ \hline
$\beta_{\mu}$ & $0.281 \pm 0.014$ & $0.494 \pm 0.04$ \\ \hline
$\mu_{e,b_j}^*$ / mag arcsec$^{-2}$ & $21.90 \pm 0.02$ & $22.82 \pm 0.19$ \\
 & ($22.45 \pm 0.02$) & \\
\hline
$\sigma_{\mu}$ & $0.517 \pm 0.012$ & $0.61 \pm 0.04$ \\ \hline
\hline
\end{tabular}
\end{table}

\begin{table*}
\caption{Table of Schechter parameters for Figure~\ref{fig:lfs} \label{table2}}
\begin{tabular}{|l|l|c|c|c|c|} \hline
$\mu_{lim}$ & Magnitude & $M^*_{b_j}-5\log\,h$ & $\phi^*/h^3$Mpc$^{-3}$ & 
$\alpha$ & $j_{b_j}/10^{8}h 
L_{\odot}$Mpc$^{-3}$ \\	\hline
\hline
26.0 & Isophotal & $-19.54\pm0.02$ & $(1.95\pm0.04)\times$10$^{-2}$ & 
$-1.057\pm0.01$ & $1.74\pm0.05$  \\  	
25.0 & Isophotal & $-19.36\pm0.02$ & $(1.85\pm0.04)\times$10$^{-2}$ & 
$-1.058\pm0.01$ & $1.40\pm0.05$  \\  
24.0 & Isophotal & $-19.10\pm0.02$ & $(1.52\pm0.04)\times$10$^{-2}$ & 
$-1.090\pm0.01$ & $0.928\pm0.05$ \\  
23.5 & Isophotal & $-18.86\pm0.02$ & $(1.36\pm0.04)\times$10$^{-2}$ & 
$-1.083\pm0.01$ & $0.662\pm0.05$ \\  
23.0 & Isophotal & $-18.66\pm0.02$ & $(1.00\pm0.04)\times$10$^{-2}$ & 
$-1.121\pm0.01$ & $0.417\pm0.05$ \\ \hline 
26.0 & Corrected & $-19.64\pm0.02$ & $(2.00\pm0.04)\times$10$^{-2}$ & 
$-1.056\pm0.01$ & $1.96\pm0.05$  \\  	
25.0 & Corrected & $-19.54\pm0.02$ & $(1.96\pm0.04)\times$10$^{-2}$ & 
$-1.055\pm0.01$ & $1.75\pm0.05$  \\  
24.0 & Corrected & $-19.34\pm0.02$ & $(1.83\pm0.04)\times$10$^{-2}$ & 
$-1.051\pm0.01$ & $1.36\pm0.05$ \\  
23.5 & Corrected & $-19.22\pm0.02$ & $(1.62\pm0.04)\times$10$^{-2}$ & 
$-1.062\pm0.01$ & $1.08\pm0.05$ \\  
23.0 & Corrected & $-19.00\pm0.02$ & $(1.43\pm0.04)\times$10$^{-2}$ & 
$-1.021\pm0.01$ & $0.760\pm0.05$ \\ \hline
26.0 & Total & $-19.74\pm0.02$ & $(2.00\pm0.04)\times$10$^{-2}$ & 
$-1.066\pm0.01$ & $2.16\pm0.05$  \\  	
25.0 & Total & $-19.74\pm0.02$ & $(2.00\pm0.04)\times$10$^{-2}$ & 
$-1.065\pm0.01$ & $2.16\pm0.05$  \\  
24.0 & Total & $-19.72\pm0.02$ & $(2.00\pm0.04)\times$10$^{-2}$ & 
$-1.058\pm0.01$ & $2.11\pm0.05$	 \\  
23.5 & Total & $-19.66\pm0.02$ & $(1.81\pm0.04)\times$10$^{-2}$ & 
$-1.037\pm0.01$ & $1.78\pm0.05$	\\  
23.0 & Total & $-19.62\pm0.02$ & $(1.08\pm0.04)\times$10$^{-2}$ & 
$-0.975\pm0.01$ & $0.989\pm0.05$ \\
\hline
\end{tabular}
\end{table*}

\appendix
\section{Testing the Face-on Parameters}
\begin{figure}
{\psfig{file=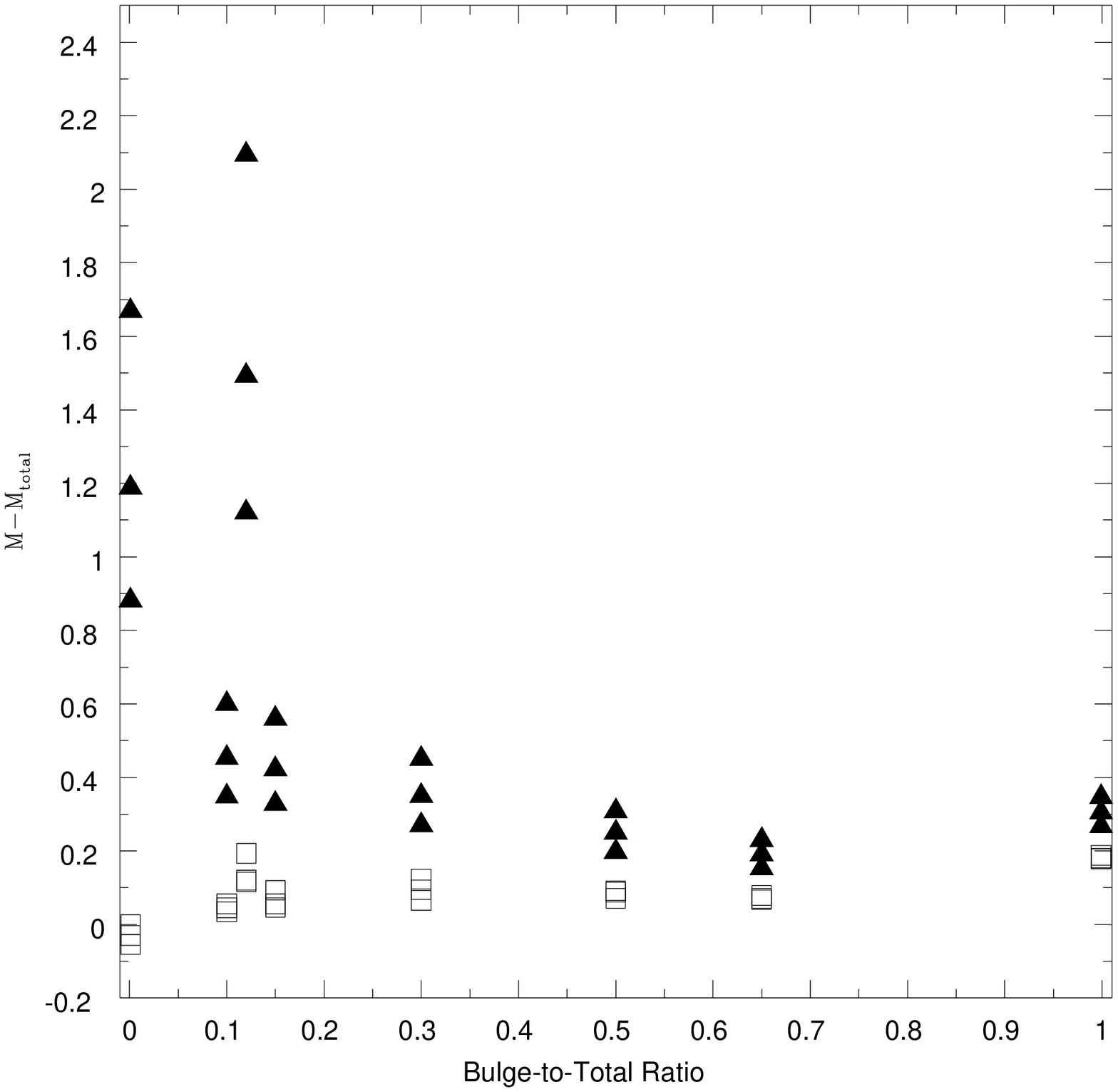,width=90mm,height=90mm}}
\caption{This plot shows the bias in corrected magnitudes and isophotal 
magnitudes as a function of bulge-to-total ratio for three different 
redshifts, $z=0.05, 0.10, 0.15$. The squares show the offset between 
exponentially corrected magnitudes (Cross et al. 2001) and total
magnitudes for an elliptical galaxy, $B/T=0.999$, an S0 galaxy, $B/T=0.65$, 
Sa-Sd galaxies, $B/T=0.5-0.1$, an irregular galaxy, $B/T=0.001$, and a LSBG 
with, $\mu_o=23$ mag arcsec$^{-2}$ and $B/T=0.12$. The offset is low generally
around 0.1 mag, with a weak dependence on redshift. The triangles show the 
difference between isophotal magnitudes and total magnitudes as a function of 
bulge-to-total ratio for the same galaxies. There is a much greater difference,
which is a stronger function of redshift. 
\label{fig:m_bias}}
\end{figure}

In Appendix A of Cross et al. (2001), we calculated the difference 
between our corrected magnitudes and total magnitudes compared to isophotal 
magnitudes and total magnitudes for 8 galaxy types. We have extended this by
taking seeing into account and looking at the errors in effective surface
brightness too. However, the galaxies are still face on. Fig~\ref{fig:m_bias}
shows the errors in magnitudes for isophotal magnitudes (triangles) and our
corrected magnitudes (squares) compared to total magnitudes, for the 2dFGRS 
data --- $\mu_{lim}=24.67$ and seeing of $2''$. Each is shown for
3 different redshifts, 0.05, 0.1 and 0.15. The corrected magnitudes are quite 
good with a typical offset of 0.1 mag and the greatest offset is 0.2 mag for 
an elliptical galaxy with a redshift of 0.1 or 0.15 and a $z=0.15$ LSBG, with a
bulge-to-ratio of 0.12. The isophotal magnitudes 
are much worse, with a typical offset of 0.4 mag and the greatest offset is 
2.1 mag for the z=0.15 LSBG. There is only a weak redshift dependence for the 
corrected magnitudes. 

\begin{figure}
{\psfig{file=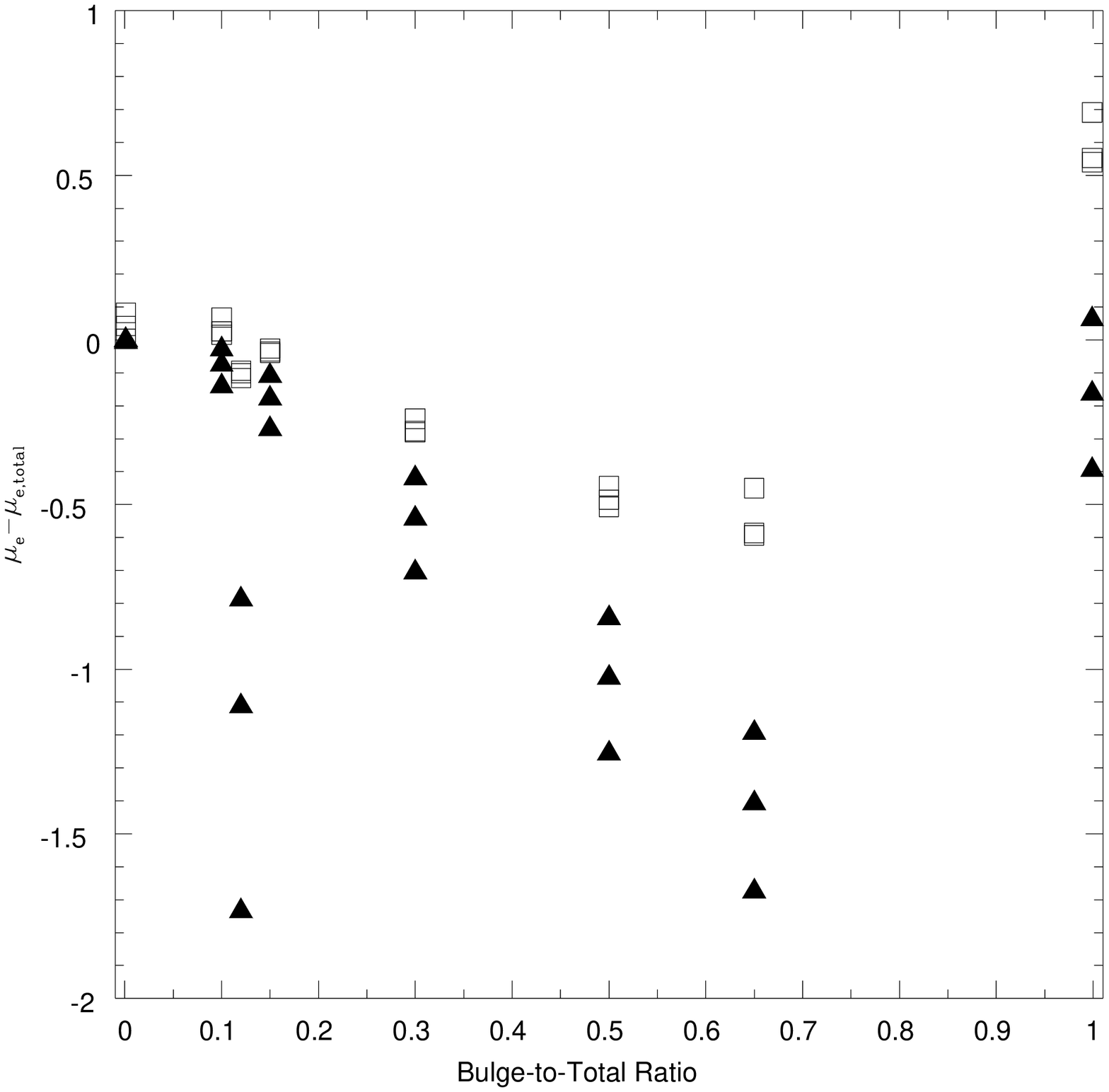,width=90mm,height=90mm}}
\caption{This plot shows the bias in effective surface brightness for 
exponentially corrected magnitudes as a function of bulge-to-total ratio. 
The galaxies are the same type as described in Fig.~\ref{fig:m_bias}. The 
triangles represent the offset when seeing is not taken into account, and the
squares show the effect of taking seeing into account. There is a weak
dependence on redshift. 
\label{fig:mu_bias}}
\end{figure}

Fig~\ref{fig:mu_bias} shows the bias in effective surface brightness for 
our correction, before and after seeing. There is up to 1.75 mag arcsec$^{-2}$
offset before seeing is taken into account and up to 0.7 mag arcsec$^{-2}$
after seeing is taken into account. The greatest offsets occur in galaxies 
with large bulges, such as Sa/S0s and 
ellipticals. The average offset will be around 0.4 mag arcsec$^{-2}$. 

However, there is a significant bias in the effective surface brightnesses. 
Ellipticals have their effective surface brightnesses underestimated by $\sim
0.5$ mag arcsec$^{-2}$; S0 and Sa galaxies have their effective surface 
brightnesses overestimated by up to 0.5 mag arcsec$^{-2}$. Late type spirals 
and irregulars will only have a negligible effect. 

However late types will be more effected by inclination. To model this, we have
assumed that a disk galaxy is optically thin and has no internal extinction. 

A galaxy of area $A_{iso}$, was assumed to have an isophotal radius $r_{iso}$,
given by:

\begin{equation}
A_{iso}=\pi\,r_{iso}^2
\end{equation}

The isophotal magnitude and radius were used to calculate the total magnitude 
and the effective surface brightness $\mu_e$ as described in Cross et al. 2001.

However a disk galaxy, inclined at an angle, $i$ with major axis $a$ and 
minor axis $b$ has an area

\begin{equation}
A_{iso}=\pi\,ab
\end{equation}

where

\begin{equation}
b=a\cos(i) 
\end{equation}

The surface brightness has increased at each point by a factor 
$\frac{1}{\cos(i)}$, increasing the semi-major axis until $\mu(a)=\mu_{lim}$.

\begin{equation}
a=\alpha\,0.4\ln(10)[\mu_{lim}-\mu_0-2.5\log_{10}cos(i)]
\end{equation}

where $\alpha$ is the disk scale length.
This increases the isophotal flux of the galaxy, as well as the isophotal 
radius. 

\begin{equation}
r_{iso}=\sqrt{ab}
\end{equation}

\begin{equation}
\begin{split}
&m_{iso}=m_{tot}-2.5\log_{10}f \\
&f=1-(1+\frac{a}{\alpha})e^{-\frac{a}{\alpha}} \\
\end{split}
\end{equation}

An exponential profile is fitted to these parameters as in Cross et al. (2001).
The central surface brightness and total magnitude are calculated for this 
galaxy assuming that the galaxy is face on. The error in the central surface 
surface brightness is the difference between the true central surface 
brightness and the calculated central surface brightness. The error in the 
effective surface brightness is exactly the same, as the difference between 
central and effective surface brightness is a constant for an exponential 
profile. The total magnitude calculated above is the same as the true total 
magnitude.

\begin{equation}
\Delta\,\mu_{0}=\mu_{0,meas}-\mu_{0,true}
\end{equation} 

The probability of a galaxy of inclination $\theta$ lying between $i$ and 
$i+di$ is:

\begin{equation}
P(i<\theta\leq\,i+di)=\frac{1}{2}\sin(i)di
\end{equation}

Therefore the mean difference in measured and face-on effective surface 
brightness is:

\begin{equation}
\bar{\Delta\,\mu_{e}}=\frac{1}{2}\int_{0}^{\pi/2}\,\Delta\,\mu_0\,
\sin(i)di=-0.477
\end{equation}

Thus the calculated effective surface brightnesses of late-type galaxies may
be 0.48 mag too bright. However more complicated effects such as seeing,
the thickness of the disk and internal extinction will all tend to reduce the
measured surface brightness of edge on disks to a greater degree than face-on
disks, reducing the mean offset. 

The overall effects of bulge-disk decomposition and inclination appear to be
to make $M^*$ brighter by 0.1 mag and $\mu_e^*$ fainter by 0.55 mag 
arcsec$^{-2}$, 0.5 mag arcsec$^{-2}$ for Sas due to the bulge and 0.1 mag 
arcsec$^{-2}$ for Sds due to the bulge and 0.5 mag arcsec$^{-2}$ due 
to inclination. It is difficult to be precise about this as the morphological 
mix of galaxies in the 2dFGRS is not known. 

\end{document}